\documentclass[namedreferences]{solarphysics}
\usepackage[hyperref,optionalrh,showbiblabels]{spr-sola-addons} 
\usepackage{graphicx}        
\graphicspath{{}}
\usepackage{natbib}

\usepackage{lineno}
\usepackage{amssymb}        
\usepackage{color}           
\usepackage{breakurl}        

\usepackage[utf8]{inputenc}  
\usepackage[T1]{fontenc}
\usepackage{bm}
\usepackage{subfig}

\usepackage{xcolor}
\usepackage{multirow}

\newcommand{\fig}{Figure~\ref}
\newcommand{\eq}{Equation~\ref}
\newcommand{\eqs}{Equations~\ref}
\newcommand{\tab}{Table~\ref}
\newcommand{\sez}{Section~\ref}

\newcommand{\quotes}[1]{``#1''}

\renewcommand{\vec}[1]{{\mathbfit #1}}

\newcommand{\pder}[2]{ \frac{\partial #1}{\partial #2} }
\newcommand{\grad}{ {\bf \nabla } }
\newcommand{\curl}{ {\bf \nabla} \times}

\newcommand{\dv}{~{\mathrm d}^3 v}

\newcommand{\ee}{\vec E}
\newcommand{\bb}{\vec B}
\newcommand{\jj}{ \vec j}

\newcommand{\xx}{ \vec x}
\newcommand{\vv}{ \vec v}
\newcommand{\kk}{ \vec k}

\newcommand{\br}{\mathbf{r}}

\chardef\us=`\_

\begin{document}

\begin{article}
\begin{opening}

\title{Statistical analysis of ions in two-dimensional plasma turbulence\\ {\it Solar Physics}}

\author[addressref={aff1},corref,email={francesco.pecora@unical.it}]{\inits{F.}\fnm{Francesco}~\lnm{Pecora}}
\author[addressref=aff2]{\fnm{Francesco}~\lnm{Pucci}}
\author[addressref=aff2]{\fnm{Giovanni}~\lnm{Lapenta}}
\author[addressref=aff3]{\fnm{{David}}~\lnm{{Burgess}}}
\author[addressref=aff1]{\fnm{Sergio}~\lnm{Servidio}}


\address[id=aff1]{Dipartimento di Fisica, Università della Calabria, Arcavacata di Rende 87036 (CS), Italy}
\address[id=aff2]{Plasma-astrophysics Section, KU Leuven, 3001 Leuven, Belgium}
\address[id=aff3]{School of Physics and Astronomy, Queen Mary University of London, 327 Mile End Road, London E1 4NS, United Kingdom}

\runningauthor{F. Pecora et al.}
\runningtitle{Statistical analysis of ions in two-dimensional simulations of plasma turbulence}

\begin{abstract}
The statistical properties of ions in two-dimensional fully developed turbulence have been compared between two different numerical algorithms. In particular, we compare Hybrid Particle In Cell (hybrid PIC with fluid electrons) and full PIC simulations, focusing on particle diffusion and acceleration phenomena. To investigate several heliospheric plasma conditions, a series of numerical simulations has been performed by varying the plasma $\beta$ - the ratio between kinetic and magnetic pressure. These numerical studies allow the exploration of different scenarios, going from the solar corona (low $\beta$) to the solar wind ($\beta \sim 1$), as well as the Earth’s magnetosheath (high $\beta$). It has been found that the two approaches compare pretty well, especially for the spectral properties of the magnetic field and the ion diffusion statistics. Small differences among the models have been found regarding the electric field behaviour at sub-ion scales and the acceleration statistics, due evidently to the more consistent treatment of the plasma in the full PIC approach.
\end{abstract}
\keywords{Plasma Physics; Turbulence}
\end{opening}

\section{Introduction}
\label{S_Introduction}

The dynamical description of astrophysical plasmas is a very challenging problem that needs to be addressed via observations, adequate theories and support from numerical simulations. In particular, spacecraft observations reveal the presence of multi-scale turbulence \citep{zimbardo10}, where kinetic physics is at work and the plasma is locally far from thermodynamic equilibrium \citep{Marsch06}. This non-Maxwellian state of plasmas suggests that self-consistent, weakly-collisional models need to be adopted \citep{Servidio15, SchekochihinEA16, Howes17}. Recent observations reveal that particles are strongly modulated by the presence of turbulence, which leads to the formation of temperature anisotropies, heat fluxes, particle beams and a wide variety of other features \citep{ServidioEA17}. These features are associated with the stochastic dynamics of particles, that cannot be observed from \textit{in situ} measurements and have to be investigated with the support of kinetic models.

Despite some restrictions in the approach (reduced dimensionality, inadequate resolution,  limitation by the available computational resources, and so on), kinetic simulations represent today a powerful tool of investigation. Although a comprehensive description of the whole turbulent cascade, which goes from the large magnetohydrodynamics (MHD) scales to the Debye scales \citep{FranciEA15,hellinger15}, is still not feasible, it is possible to describe this process in a limited range of wavelengths, especially at ion and sub-ion scales. In this case, the Vlasov--Maxwell system of equations needs to be solved - a difficult task that can be tackled via two main approaches. The first is based on the Eulerian approach, with which the equations are solved over a multi-dimensional grid \citep{Valentini07}  - a method which is quite time-consuming and memory-demanding. The second is the PIC philosophy, which uses a Lagrangian approach to follow the dynamics of so-called {\it macro-particles}, intended as \quotes{slices} of the initial velocity distribution function (VDF) \citep{Markidis10}. This approach is faster and can be used to describe three-dimensional (3D) systems.  The drawback is that, because of the VDF sampling of the PIC codes, the number of the computational particles is relatively small and a noise background is introduced by these finite-size, pseudo-particles. 

In the PIC framework, there are two main ways to model the plasma, either via the hybrid (I) or the full kinetic (II) description. Hybrid codes are useful when the scales are in between the MHD range and the sub-ion lengths. With this description ions and electrons are treated differently: the former are present as particles whereas the latter are modelled as a massless fluid that neutralises the plasma and provides a pressure term \citep{Winske85}. This separation of scales retains the ion kinetic effects at the price of neglecting those of the electrons, granting a reduced computational cost \citep{Matthews94}. The approach (II) is represented by full kinetic schemes, where both ions and electrons are treated via the full kinetic Vlasov models and it is (obviously) more expensive from the computational point of view.

In this work, we compare hybrid PIC (I) and full PIC (II) codes, by varying numerical parameters, such as the resolution and the number of particles per cell, as well as the physical settings.  We varied the plasma $\beta$ to see how particles behave in different turbulent scenarios, characterising different environments such as the solar corona $(\beta \ll 1)$, the solar wind $(\beta \sim 1)$, and magnetosheath conditions $(\beta > 1)$. 

The simulations are performed in a 2D spatial plane with a mean magnetic field in the out-of-plane direction. The fields have three components but depend only on the two in-plane coordinates. This geometry is useful when describing plasmas with a strong mean magnetic component such as coronal loops \citep{Einaudi96}, the solar wind \citep{bieber962DSW} and also plasma fusion devices \citep{ongena16}.

It is important to note that more technical parameters may also influence the results. For instance, an inadequate resolution might affect energization phenomena as well as the diffusion of particles. Analogously, the number of particles per cell (ppc) is of fundamental importance since an excess of noise can lead to numerical collisionality and a consequent fictitious heating \citep{kim05}. In this work, we will vary both the above numerical parameters.

This work is organised as follows. In \sez{sec:Codes} we give an overview of the two numerical approaches used, along with their general settings. In \sez{sec:Results} we show the results focusing on the energy conservation (\ref{sec:En_cons}), electric and magnetic field power spectra (\ref{sec:spectra}), the ion motion (\ref{sec:diff}), and the plasma heating (\ref{sec:heating}). Finally, the discussions and the main conclusions will be presented in the last section.

\section{Plasma Turbulence Simulations}
\label{sec:Codes}
In this section we give an overview of the simulations, focusing on the description of the two different approaches and the numerical parameters. We use a 2.5-dimensional (2.5D) geometry, which means that the fields depend on two spatial components but they have all the three vector components. A mean field is imposed along the out-of-plane direction ${\mathbf{B}}_0 = B_0\hat{\mathbf{z}}$. It has been already shown that turbulence becomes anisotropic when a mean guiding field is present and it develops essentially in the plane perpendicular to the mean field \citep{Shebalin83, Matthaeus86, Dmitruk04}. However, a limitation of this geometry is the neglect of parallel propagating waves and compressive fluctuations along the mean magnetic field. A fully 3D simulation with a large range of scales that retains all these effects is much more computationally expensive and it will be investigated in future works.

In our simulations we use three values of plasma $\beta = 0.1, 0.5, 5$, for each species. All the simulations have identical initial conditions, with uniform density and temperature, and the VDFs are Maxwellian with different bulk speeds. The initial fluctuations correspond to a superimposition of modes in the plane perpendicular to $\mathbf{B}_0$, with zero initial parallel variances. The fluctuations amplitude is $\delta b/B_0 \sim 0.3$. The physical size of the box is $L_x \times L_y= \left( 128\times 128 \right) d_i^2$ with periodic boundary conditions.

\subsection{Hybrid PIC simulations}
\label{sec:hybrid}
As mentioned earlier, the hybrid PIC simulations were performed at three different values of the plasma $\beta$ (equal for both protons and electrons), defined as $\beta \equiv \beta_e = \beta_i = P_{kin}/P_B = 2 v_{th}^2/v_A^2$, with $v_{th}$ and $v_A$ being respectively the initial thermal speed and Alfv\'en speed (related to the mean magnetic field $B_0$). For both hybrid and full PIC simulations, particles and fields are coupled via the Vlasov--Maxwell equations. For the hybrid approach, a generalized Ohm's law for the electric field is given. The equations read as:
\begin{equation} 
\def\arraystretch{2.5}
\begin{array}{r@{}l}
\dot{\xx} &{}= \vv,\\
\dot{\vv} &{}= \ee + \vv \times \bb,\\
\displaystyle \pder{\bb}{t} &{}=-\curl \ee=\curl\left(\vec{u} \times \bb - \frac{1}{n} \; \jj \times \bb + \frac{1}{n} \grad P_e - \eta \jj \right),
\end{array}
\label{eq:hybridPIC}
\end{equation}
where $\xx$ is the particle's position and $\vv$ its velocity. $\bb$ and $\ee$ represent the magnetic and electric fields respectively, $\jj = \curl \bb$ is the current density. The ion number density is $n=\int f \dv$ and $\vec u = (1/n)\int \vv f \dv$ is the bulk velocity. The ion VDF is $f(\xx, \vv, t)=f(x, y, v_x, v_y, v_z, t)$. The electron pressure term is given by an adiabatic equation of state $P_e = \beta n^\gamma$, with $\gamma=5/3$ being the classical adiabatic index, while the resistive term $\eta \jj$, with $\eta=0.006$, is introduced to grant dissipation at small scales and prevent numerical instabilities. Lengths are normalized to the ion skin depth $d_i = c/\omega_{p_i}$, where $c$ is the speed of light and $\omega_{p_i}$ is the ion plasma frequency. Time is normalised to the inverse of ion cyclotron frequency $\Omega_{ci}^{-1}$, velocities to the Alfv\'en speed $v_{{}_A}= c \Omega_{ci}/\omega_{p_i}$. To simulate the motion of the big energy-containing vortices we impose random fluctuations at large scales for both the magnetic field and the protons' bulk velocity flow, as described before.

As already stated, to ensure a low noise value, implicitly coming from the PIC method, we use $1500$ particles per cell. The details are summarised in \tab{tab:Hruns}. More details on these simulations are reported in previous works \citep{Servidio16,pecora18}.

\begin{table}
\begin{tabular}{ ccccc }
ID & $N_x \times N_y$ & ppc & $\beta$ \\
\hline
H1 & $512 \times 512$ & 1500 & 0.1\\
H2 & $512 \times 512$ & 1500 & 0.5\\
H3 & $512 \times 512$ & 1500 & 5\\
\hline
\end{tabular}
\caption{Parameters of the hybrid PIC simulations. The first column is the label identifying the simulation, $N_x$ and $N_y$ are the number of cells in the $x$ and $y$ directions respectively, ppc is the number of particles per cell, and in the last column we report the plasma $\beta$.}
\label{tab:Hruns}
\end{table}

\subsection{IPIC full PIC simulations}
\label{sec:pic}
The full kinetic simulations were performed with an implicit PIC method implemented in a 3D parallel code, called iPIC3D \citep{Markidis10}. The simulations were carried out using the same values of the ion $\beta$ as in \sez{sec:hybrid}. The governing equations in the full kinetic simulation, in code units, are given by
\begin{equation}
\def\arraystretch{2.5}
\begin{array}{r@{}l}
&{}\displaystyle\frac{\partial f_s}{\partial t} + \vv \cdot \frac{\partial f_s}{\partial \xx} + \frac{q_s}{m_s}\left( \ee +  \vv \times \bb \right) \cdot \frac{\partial f_s}{\partial \vv} = 0,\\
&{} \displaystyle \grad ^2 \ee - \frac{\partial ^2 \ee}{\partial t^2} = \displaystyle 4\pi \pder{\jj}{t} + 4\pi\grad \rho,\\
&{}\displaystyle \pder{\bb}{t} = - \curl \ee.
\end{array}
\label{eq:VM}
\end{equation}
In this case, lengths are normalized to $d_i$, times to $d_i/c$ and velocities to $c$, that is set to $1$ in code units. The mass ratio used is $m_i/m_e = 25$, and $v_{{}_A}/c = B_0 = 10^{-2}$. $(q_s/m_s)$ is the charge-to-mass ratio of the species $s$, normalized to the physical ion charge-to-mass ratio. $\omega_{pe}/\Omega_{ce} = \sqrt{m_e/m_i}/\Omega_{ci}$. $\rho = \sum_s q_s \int f_s \dv$ and $\jj = q_s \int \vv f_s \dv$ are the charge density and the current density computed over the two species $s$. The set of \eqs{eq:VM} are solved over the same physical domain as in Sec.\ref{sec:hybrid}, whereas the discretization varies as reported in \tab{tab:Kruns}. The electric and magnetic fields are given by the Maxwell equations whose solution is computed implicitly, meaning that, with respect to a time step $n$, the charge density $\rho$ is evaluated at time $n+1$ and the current density $\jj$ at an intermediate step $n+1/2$ \citep{Markidis10}. The time step is $\Delta t = (1/8\pi) \, \tau_{ge}$, with $\tau_{ge} = 2\pi/\Omega_{ce}$ being the electron cyclotron frequency.

\begin{table}
\begin{tabular}{ cccccc }
ID & $N_x \times N_y$ & ppc & $\beta_e = \beta_i$ & $\beta_{tot}$ & $\Psi_{max}$\\
\hline
run1 & $512 \times 512$ & 400 & 0.1 & 0.2 & -1.8\% \\
run2 (K1) & $512 \times 512$ & 4000 & 0.1 & 0.2 & -1.4\%\\
run3 & $1024 \times 1024$ & 1000 & 0.1 & 0.2 & -1.2\%\\
\hline
run4 & $512 \times 512$ & 400 & 0.5 & 1 & -3.0\%\\
run5 (K2) & $512 \times 512$ & 4000 & 0.5 & 1 & -0.3\%\\
run6 & $1024 \times 1024$ & 1000 & 0.5 & 1 & -2.5\%\\
\hline
run7 & $512 \times 512$ & 400 & 5 & 10 & -13.5\%\\
run8 (K3) & $512 \times 512$ & 4000 & 5 & 10 & -1.9\%\\
run9 & $1024 \times 1024$ & 1000 & 5 & 10 & -9.6\%\\
\hline
\end{tabular}
\caption{Parameters used in the full PIC code. ID is the label identifying the simulation, $N_x$ and $N_y$ are the number of cells in the $x$ and $y$ directions respectively, ppc is the number of particles per cell, $\beta_e$ and $\beta_i$ are the plasma $\beta$ for electrons and ions respectively and $\beta_{tot}$ is the total plasma $\beta = \beta_i + \beta_e$. $\Psi_{max}$ is the maximum energy variation of the run as defined in \eq{eq:psi} and depicted in \fig{fig:K_en_cons}.}
\label{tab:Kruns}
\end{table}

\section{Results}
\label{sec:Results}
In this section, we present the results from the comparison between the above two approaches. First, we look at the main differences among the full PIC simulations to see whether the numerical parameters affect the general behaviour. After a first overview, we chose the \quotes{best} simulation for each $\beta$ and compare it with the corresponding hybrid PIC case.

\subsection{Energy Conservation}
\label{sec:En_cons}
One of the major issues that arises when dealing with simulations is the total energy conservation. We define a measure of the conservation of the energy, as

\begin{equation}
\Psi(t) = \frac{\mathcal{E}(t) - \mathcal{E}(0)}{\mathcal{E}(0)},
\label{eq:psi}
\end{equation}
where $\mathcal{E}$ is the total plasma energy. The energy is given by magnetic $\mathcal{E}_B = \int d\br B^2/8\pi$ and electric $\mathcal{E}_E = \int d\br E^2/8\pi$ contributions, and the particles total kinetic energy $\mathcal{E}_K = 1/2 \sum_\alpha \sum_p m_\alpha v^2_{\alpha,p}$, where the index $p$ runs over all the particles and $\alpha$ over the two species (ions and electrons).

\begin{figure}
\centerline{
\includegraphics[width=.7\textwidth]{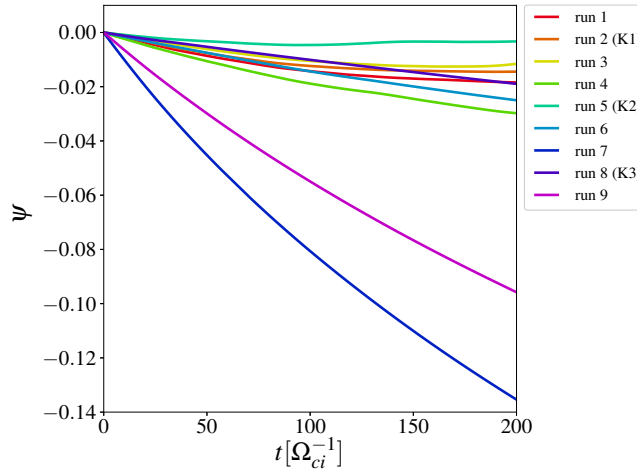}
}
\caption{Total energy conservation for the full PIC simulations. All of them conserve energy within $3\%$ of the initial value but run 7 and run 9 go down to $\sim 14\%$ and $ \sim 10\%$, respectively.}
\label{fig:K_en_cons}
\end{figure}

\fig{fig:K_en_cons} shows the energy conservation parameter $\Psi$, for all the full PIC simulations. The general behaviour is very good since eight runs out of ten do conserve energy within $\sim 3\%$ of the initial value. All the losses are due to the semi-implicit method. Only two runs go down to $\sim 10\%$ and $\sim 14\%$. In these latter two cases $\beta_{tot} = 10$, suggesting that less magnetized plasmas need to be treated more carefully. In fact, higher $\beta$ means larger excursions in the velocity subspace, and therefore the number of particles should increase to correctly reconstruct the VDF. This means that the wider the distribution (the less magnetized the plasma), the more particles are needed to sample the distribution accurately. As it can be seen from \fig{fig:K_en_cons}, run 7 is the worst simulation regarding energy conservation and doubling both the resolution and the number of particles does not imply a noticeable improvement (run 9). Instead, a huge improvement in energy conservation is achieved when the number of particles per cell is increased by one order of magnitude even with less resolution (run 8). When the plasma $\beta$ is average or low, the energy conservation is excellent and quite similar among all the runs.

The best energy conservation is achieved in run 5 - large number of ppc and low $\beta$. This leads us straightforwardly to the choice of run 5 and run 8 as the best candidates to look at the physical effects, along with run 2. These full PIC runs are going to be compared with the hybrid PIC H1 ($\beta=0.1$), H2 ($\beta=0.5$) and H3 ($\beta=5$).

\subsection{Power Spectra}
\label{sec:spectra}
\begin{figure}
\centering
\includegraphics[width=.75\textwidth]{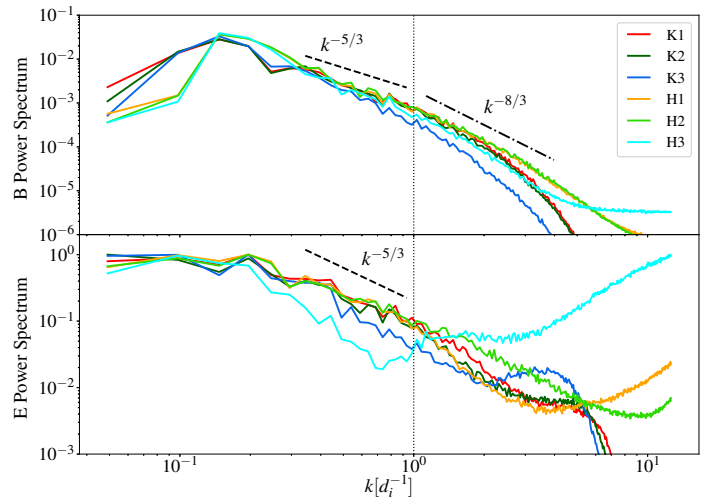} \\
\caption{Comparison of the magnetic field power spectra for the hybrid (H1, H2, H3) and the full PIC (K1, K2, K3) simulations. In both cases, the spectra follow the Kolmogorov's prediction $k^{-5/3}$ in the inertial range ($k$s smaller than the ion skin depth) and the magnetic field spectrum approaches a $k^{-8/3}$ power law at smaller scales.}
\label{fig:HK_EB_spectra}
\end{figure}

Both the hybrid and the full kinetic simulations show important features in the power spectra of magnetic and electric fields, as shown in \fig{fig:HK_EB_spectra}. The magnetic field spectrum manifests, for scales larger than the ion skin depth, an inertial range consistent with the Kolmogorov prediction of fluid-like turbulence, where the power spectrum scales as $k^{-5/3}$. At smaller scales, the spectral slope is steeper, consistent with the index $-8/3$, typical manifestation of dispersive-kinetic physics \citep{Alexandrova09,FranciEA15}. The spectra for the magnetic field compare quite well for two methods in the inertial range, but there are some small differences at sub-ion scales. In particular, in the very high-$\beta$ plasma, the magnetic field is steeper for the full PIC simulation. This may be due to electron physics, where electron resonances and damping might interact with small-scale magnetic fluctuations. Note also that for the hybrid case, at very small scales (almost the grid size), there is a flattening of the spectra, due to particle noise which has not been treated with numerical filters (see below). This is indeed more evident for the high-$\beta$ simulation, H3, at $k d_i>2$.  The different behaviour between the two codes at the very small scales is therefore due to the different ways of handling the noise. In the Hybrid simulation there is no artificial small-scale filter for the noise. In the full PIC algorithm, a local smoothing technique has been introduced, as described in previous works \citep{Olshevsky18}.

To better highlight small scale differences between the hybrid and the full PIC approach, we also computed the power spectrum of the electric field, shown in \fig{fig:HK_EB_spectra} (bottom panel). Qualitatively, these spectra are all in agreement with spacecraft observations \citep{Bale05}, for which the slopes of the two fields are similar in the inertial range, whereas at sub-ion scales the electric field spectrum is higher than that of the magnetic field. There are some small differences between the two numerical methods, which are due to both physical and numerical reasons. Regarding the numerical reason, the noise kicks in, in the hybrid case, at $k d_i>2$, as discussed before. The particle noise is more evident now since the electric field is more sensitive to small scale fluctuations and more directly to the noise of the particles in the momentum, since $\bm E \sim \bm u\times \bm B$. Regarding the physical reasons, at scales close to (and smaller than) the ion skin depth, it is interesting to note that, for the low $\beta$, there is more electric power in the full PIC case. This is very likely due to the fact that, generalising Ohm's law, many contributions are missing in the hybrid approximation with respect to the full PIC case (such as the divergence of the whole electron pressure tensor contribution and other smaller electron inertial terms).
Statistical convergence has been verified for the hybrid run \citep{Servidio16} and the full PIC. In the latter case, which is very sensitive to particle noise, we did a convergence study (shown in the Appendix for the high $\beta$ case that is the most sensitive to noise) in which we found that: (i) the power spectrum of the magnetic field is consistent going from 400 ppc to 4000 ppc, and (ii) that the diffusion coefficient is consistent between these cases with different models and resolutions. The latter is not surprising since the diffusion coefficient depends mostly on energy-containing and inertial scales of the magnetic spectrum ($k \ll d_i^{-1}$). In the highest $\beta$ case, for our best simulation, the error for energy conservation is about $1-2\%$.

\subsection{Ion Diffusion}
\label{sec:diff}
\begin{figure}[ht]
	\centering
	{\includegraphics[width=.45\textwidth]{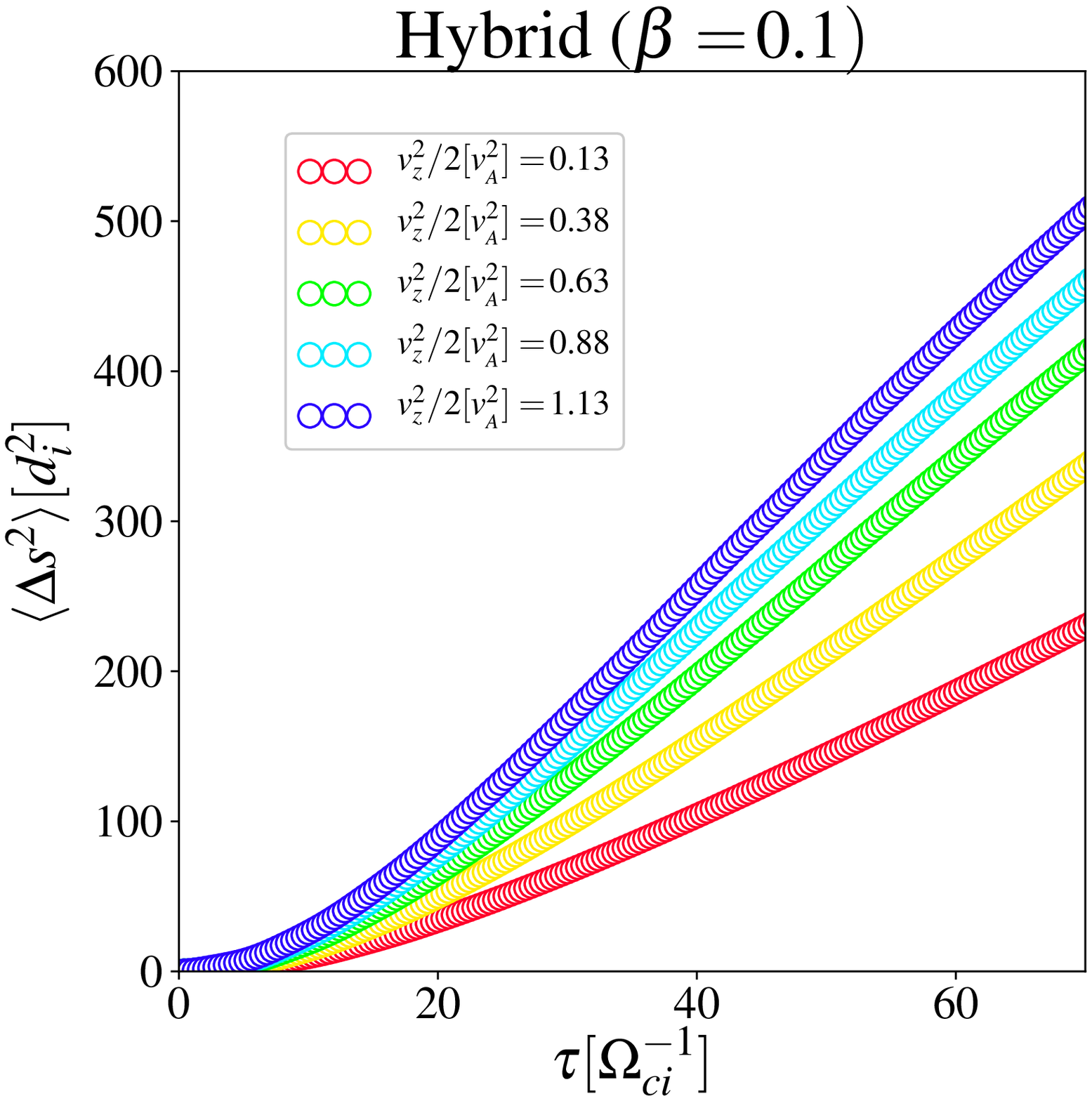}}
  \hspace{0.5cm}
	{\includegraphics[width=.45\textwidth]{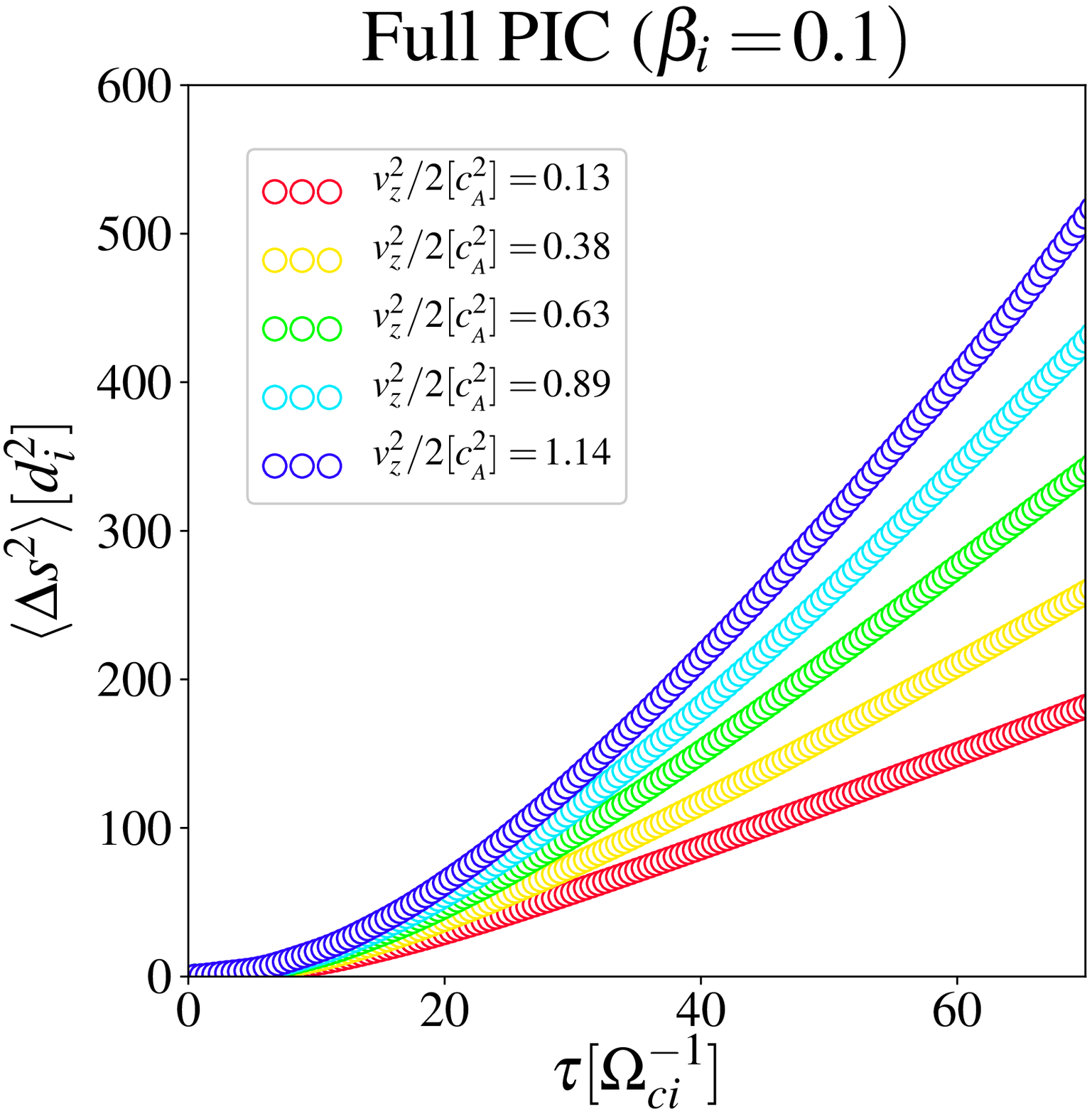}}

	{\includegraphics[width=.45\textwidth]{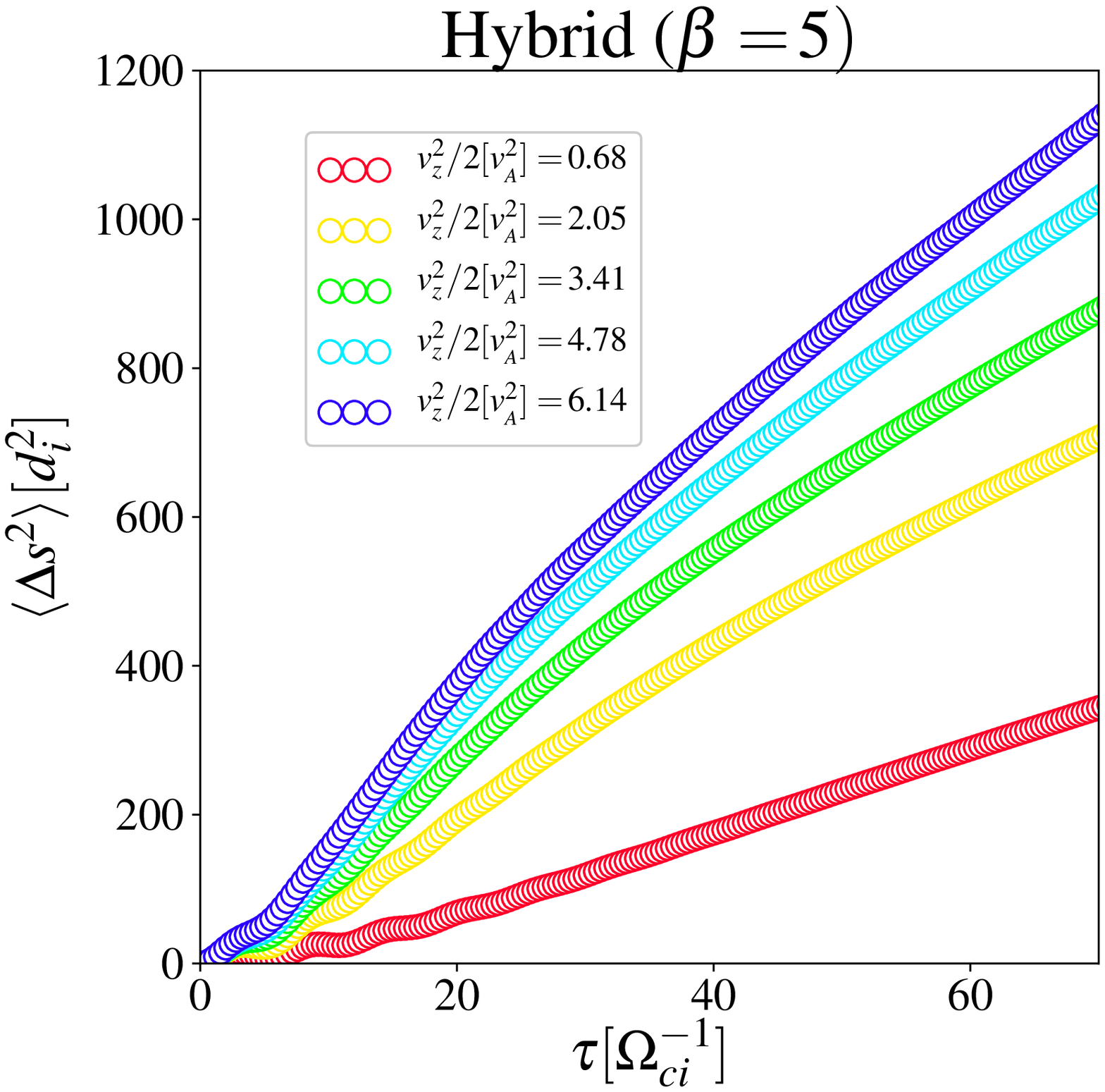}}
  \hspace{0.5cm}
  	{\includegraphics[width=.45\textwidth]{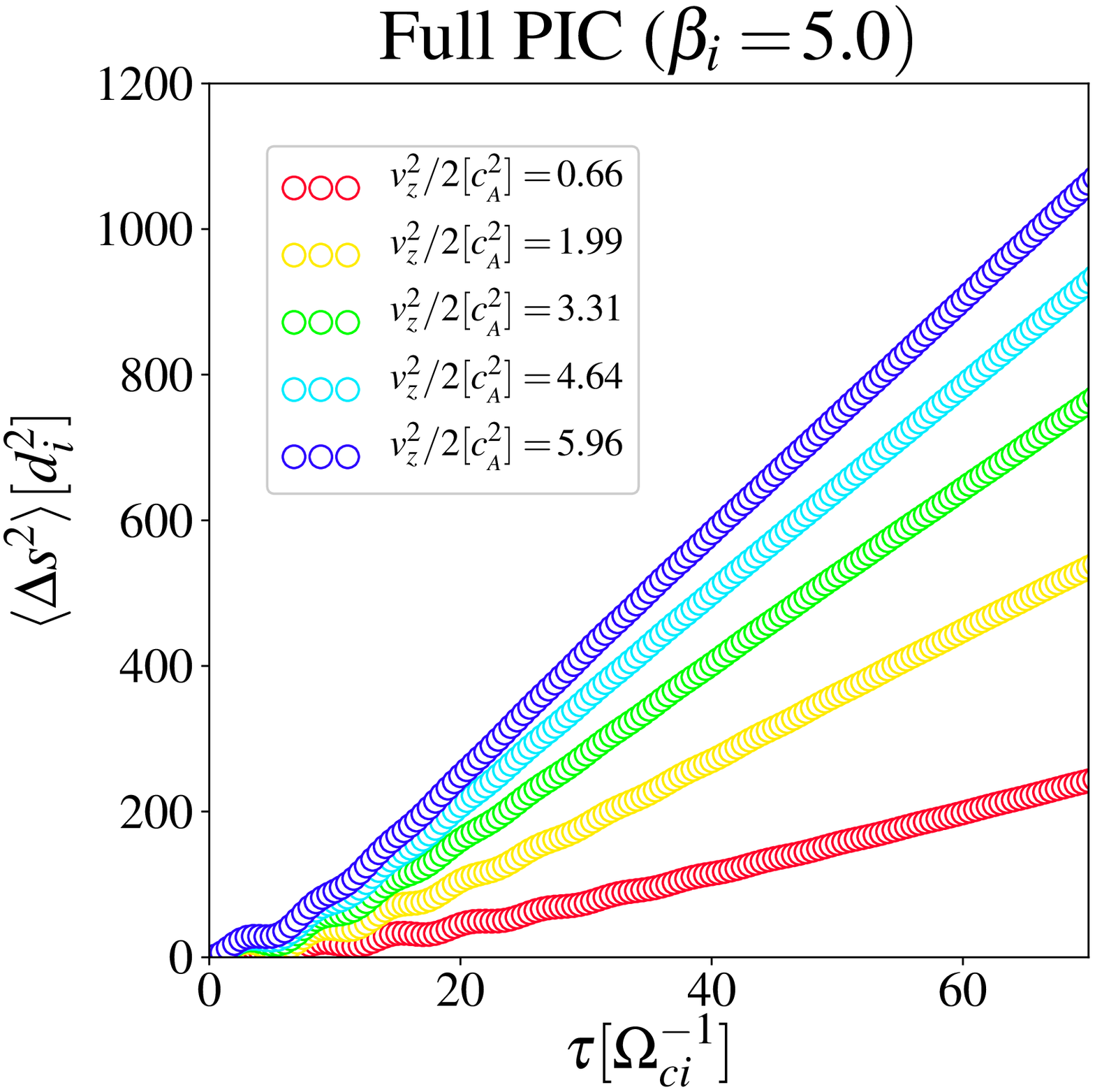}}
  
   \vspace{0.5cm}
   \caption{Ion mean squared displacement for the hybrid and the full kinetic simulations at the lowest and the highest values of the plasma $\beta$. For each $\beta$ the particles energy class value is reported in the label.}\label{fig:run_Ds2}
\end{figure}

This section is focused on the motion of the macro-particles. To have a look at the collective general motion of the ions we computed the mean squared displacement in the $x-y$ plane, namely $\Delta s^2 = \Delta x^2 + \Delta y^2$ and $\Delta x$ and $\Delta y$ are the particle's displacements in the $x$ and $y$ directions respectively, defined as $\Delta x = x(t_0+\tau)-x(t_0)$ and $\Delta y = y(t_0+\tau)-y(t_0)$ and $\tau$ is the time interval over which the displacement is calculated. If the motion is diffusive, this displacement can be described as \citep{Chandra43R,batchelor1976brownian,wang2012brownian}
\begin{equation}
\langle \Delta s ^2 \rangle = 2 D \tau,
\label{eq:diff}
\end{equation}
where $D$ is the diffusion coefficient. The measured mean squared displacements are shown in \fig{fig:run_Ds2}, only for a few energy classes, for the sake of clarity.

In all cases, the linear trend expected for the diffusion is achieved after long time intervals. The partitioning of particles in parallel energy $(v_z^2/2)$ bins has been performed to control whether it influences perpendicular diffusion as predicted by the 2D Non-Linear Guiding Center theory (2D-NLGC) \citep{Matthaeus03,Ruffolo12,pecora18}. The diffusion coefficient predicted by the theory is
\begin{equation}
    D^* \sim \sqrt{\frac{v_z^2}{B_0^2} \int d\kk \frac{S(\kk)}{k^2}},
    \label{eq:Dstar}
\end{equation}
where $\kk$ is the wave vector and $S(\kk)$ is the power spectrum of the magnetic field. Note that in Equation \ref{eq:Dstar} we used the approximation where the time decorrelation does not enter the dynamics, namely when the field can be considered time-independent, as discussed in \citet{pecora18}. As expected from the theory, the diffusion coefficient is somehow proportional to the energy of the particles in the parallel direction and to the presence of magnetic turbulence.

\begin{figure}[ht]
	\centering
  	{\includegraphics[width=.6\textwidth]{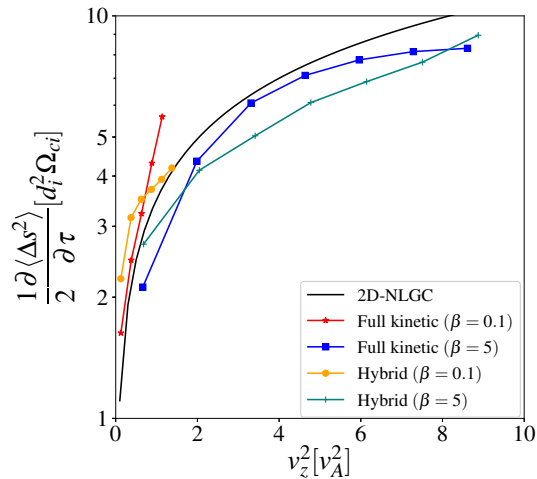}}
   \vspace{0.5cm}
   \caption{Perpendicular diffusion coefficient measured for the lower and higher $\beta$ values, for the hybrid and full kinetic runs. The \textit{solid (black) line} represent the theoretical prediction in \eq{eq:Dstar}}.
   \label{fig:Dperp_2DNLGC}
\end{figure}

To study the process of diffusion in our simulations, we computed the so-called {\it running diffusion coefficient}, defined as
\begin{equation}
    D = \frac{1}{2}\pder{\langle \Delta s^2 \rangle}{\tau}, 
\end{equation}
for each energy class. The above quantity, computed over an ensemble of particles, achieves a plateau after the diffusive limit is reached, as reported in \citet{Servidio16}. This classical procedure gives a measure of the diffusion coefficients for each class of parallel energy. The measured values are reported in \fig{fig:Dperp_2DNLGC}, for $\beta =0.1 \mbox{ and } 5$, for both types of numerical experiments. A quite good agreement between both numerical approaches and the theory is found.

\subsection{Particle Heating}
\label{sec:heating}
\begin{figure}[ht]
	\centering
	{\includegraphics[width=.45\textwidth]{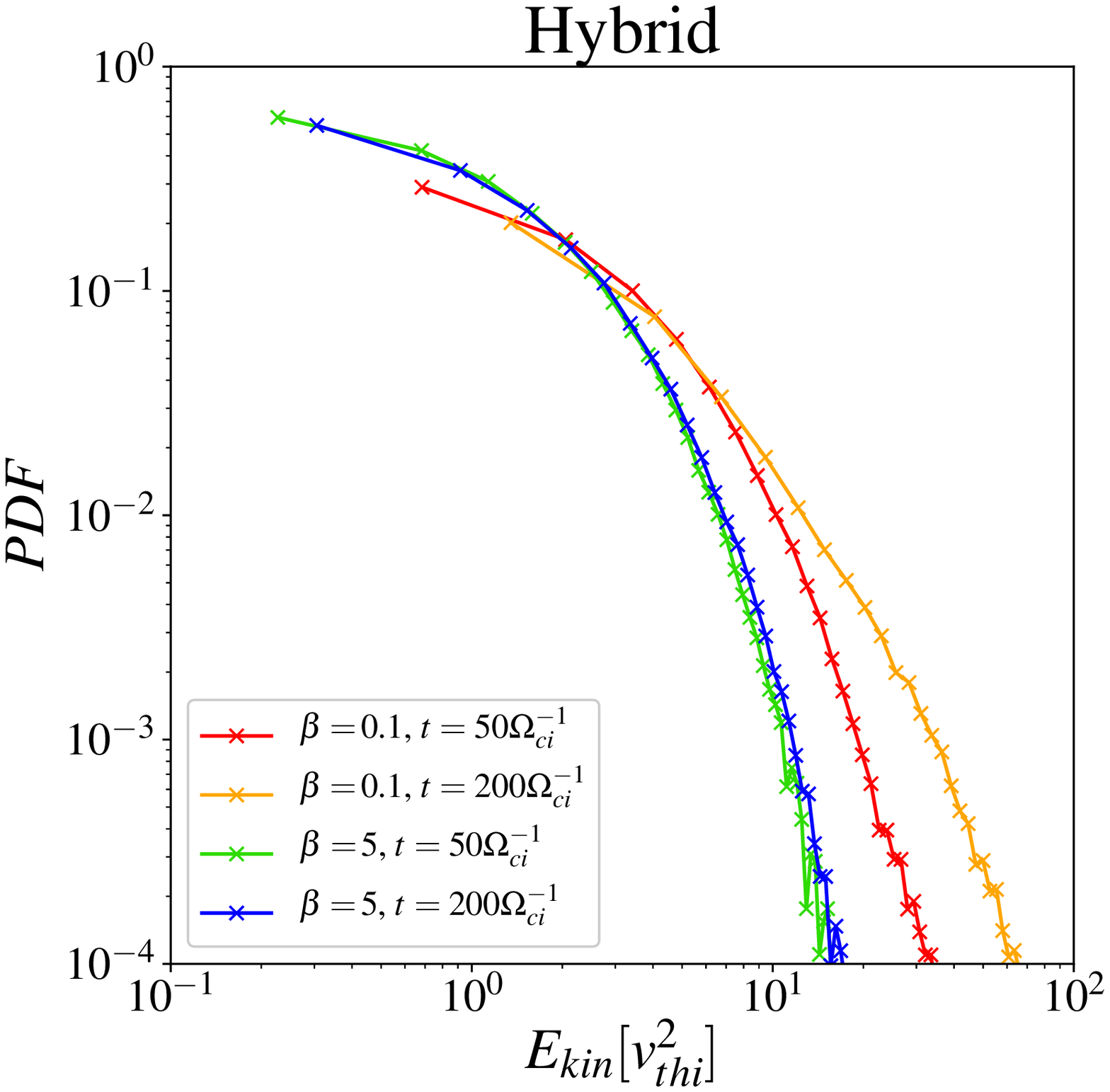}
	\label{fig:Ekin_H}}
  \hspace{0.5cm}
	{\includegraphics[width=.45\textwidth]{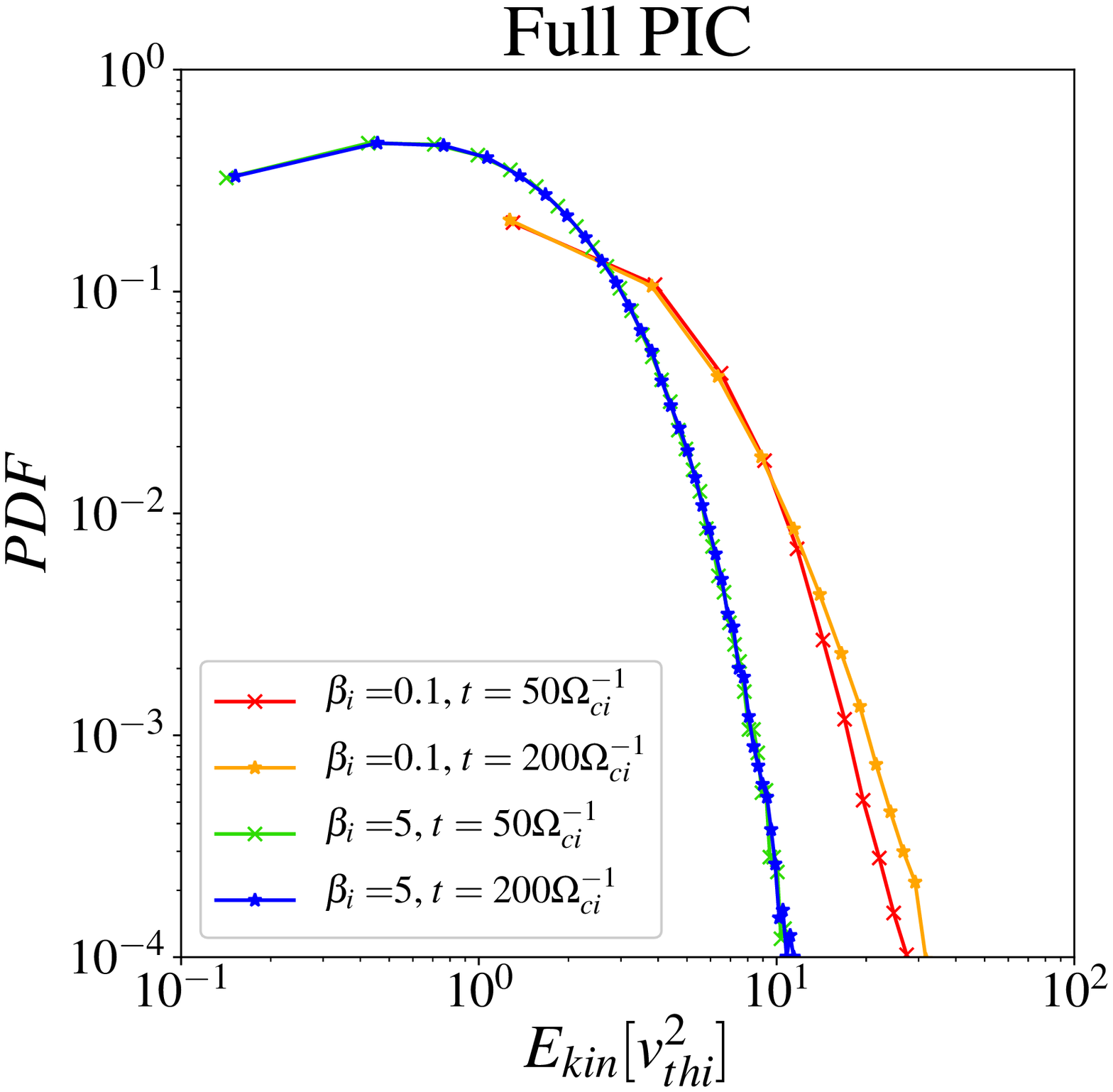}
	\label{fig:Ekin_K}}
  \hspace{0.5cm}
   \vspace{0.5cm}
   \caption{Particles kinetic energy $E_{kin}$ evolution when turbulence reaches its peak, and at the final time. The energy is normalized to the ion thermal velocity $v_{thi}$}
   \label{fig:Ekin}
\end{figure}

Along with diffusion, the process of heating is currently one of the major topics in space plasma physics. It is important to establish whether there are substantial differences for the particle energization mechanisms, comparing the hybrid PIC and of the full PIC simulations. The kinetic energy of the particles has been measured as $E_{kin} = (v_x^2 + v_y^2 + v_z^2)/2$ at the beginning of the turbulence steady state $(t\Omega_{ci} = 50)$ and at the final time of the simulations $(t\Omega_{ci} = 200)$. The probability density function (PDF) of the particles kinetic energy is shown in \fig{fig:Ekin}.

The PDF comparison suggests that in the hybrid case, at high $\beta$, there is a very small deviation between the two times, indicating a lack of energisation. This is because of particles, in the case of large $\beta$ (and large Larmor radius), do not gain energy effectively by interacting with the turbulence structures such as current sheets \citep{pecora18}. This scenario changes when one looks at the low $\beta$ case, where an extended tail develops at later times. This confirms the picture of small energy particles being able to actively interact with current sheets because their scales are comparable, and undergo a stochastic acceleration mechanism \citep{ChandranEA10}. This view is almost maintained in the full kinetic simulations, with small differences. In the low-$\beta$ case, the tail is slightly less pronounced indicating that, when electrons are also treated kinetically, the ions gain less energy from the turbulent fields. This suggests that, very likely, the electrons participate more effectively to this turbulence--particle interaction. Electrons may interact more synergically with the structures generated by turbulence, such as vortices, current sheets and from reconnection events \citep{DrakeEA10,HaynesEA14}. This latter phenomenon will be a matter of future investigations.

\section{Conclusions}     
     
In this work, we compared two of the most commonly used approaches for plasma physics simulations: (I) the hybrid and (II) the full kinetic PIC algorithms. We used the same initial conditions to highlight possible differences that could arise during the evolution of the simulations. We presented simulations at three different values of the plasma $\beta$, in order to describe a number of scenarios for heliospheric plasmas, going from the solar atmosphere ($\beta\lesssim 0.2$) to the outer regions such as the fast/slow solar wind ($\beta \simeq 1$) and the turbulent magnetosheath ($\beta\gg 1$), and by varying also numerical parameters such as the number of ppc and the meshes.  We performed several simulations, for each of the three $\beta$s, using the full kinetic code with different numerical parameters, investigating how the energy conservation is affected by the choice of the numerical setup. We found that the resolution (the number of cells used to discretize the computational domain) is of secondary importance with respect to the number of ppc, especially for energy conservation.

The analysis of the power spectra of turbulence revealed that the magnetic and electric field spectra are consistent with the Kolmogorov  prediction in the inertial range. The magnetic field spectrum exhibits a steeper power law at smaller (kinetic) scales, with $S(k)\sim k^{-8/3}$. The results were comparable within the two codes.  Regarding the electric field, the differences between the two simulations are more appreciable. At smaller scales $( k \geq d_i^{-1})$ and at low $\beta$, the electric field spectrum of the full PIC simulation is larger than that of the corresponding hybrid simulation. This is probably due to the fact that in the full kinetic case the electric field retains the contribution of small scale effects, such as the electron pressure-divergence term.

In the final part of our comparison, we focused on the processes of diffusion and energization. To study the diffusive process, we measured the mean squared displacement of each particle and averaged it over ensembles of particles belonging to the same energy range. We were able to see that the ions, in both cases, and at each $\beta$, reach a diffusive behaviour after roughly $70 \Omega_{ci}^{-1}$. The comparison between the two approaches is quite good, indicating that ions follow a modified nonlinear guiding centre model of diffusion, here adapted to the 2D case. The similarity between the hybrid and the full PIC code is not surprising, since diffusion is governed more by large, energy-containing scales, and is less sensitive to micro-physics.

The process of particle energization is similar between the two numerical approaches: ions are better energized at lower $\beta$s, even though it is slightly more evident in the hybrid PIC approach. This is due to a process of resonance with current sheets, as described in literature \citep{ChandranEA10,DrakeEA10,pecora18}. The presence of non-ideal electric field can enhance the acceleration process, as shown in previous works \citep{comisso2018}, and is more effective at low $\beta$ \citep{ball2018}. It is important to notice that, in the full PIC code, other mechanisms might preferentially heat the electrons rather than the ions, even though the unphysical mass ratio used here cannot clarify completely the possible competition taking place between the two species \citep{Daughton11}. We, therefore, restricted our comparison to the ion and sub-ion scales and will treat the sub-electron scales in future works.
  
The obtained results have several implications for the understanding of the solar wind  \citep{BrunoCarbone16}, suggesting that turbulence might enhance the process of plasma heating \citep{Verma1995}. In this complex, collisionless environment, the interaction between particles and coherent structures such as magnetic islands, reconnecting regions and waves patterns, play a fundamental role \citep{Marsch06,Kasper2008,Osman10}. The turbulent heating might help to explain the larger than expected temperature of the solar wind as it radially expands through the heliosphere \citep{Smith01}, as well as the diffusion of solar energetic particles in the heliosphere \citep{Jokipii77,Isenberg05,Tessein13}. These results are also particularly relevant for recent observations of non-Maxwellian velocity distribution functions in the turbulent magnetosheath \citep{Burch16,ServidioEA17,Yamada18}, where it has been suggested that the interaction with coherent structures produces non-thermal features. Particle diffusion due to turbulence is particularly relevant in the higher corona, where turbulence enhances the diffusivity of plasma elements, as observed for cometary-tail particles \citep{DeForest15}. The implications of these results on the understanding the pristine solar wind are particularly important for recent solar space missions such as \textit{Parker Solar Probe} \citep{Fox16PSP} and the upcoming \textit{Solar Orbiter} mission.

In future works, we plan to study the case in which the mean field is in the plane, or has an in-plane component, as discussed in several pioneering works \citep{Ofman07,Ofman10,Maneva13,Ofman14,Maneva15,Ofman17}. We also plan to compare hybrid and full particles in 3D. The 2.5D assumption might affect the power spectrum of the magnetic fluctuations, in particular at (and beyond) the typical proton scales, especially for high $\beta$. However, our geometry can describe the most intense component of turbulence, namely in the plane perpendicular to the mean field \citep{Shebalin11}.  It has been suggested, indeed, that these reduced models of collisionless plasmas share several similarities with full 3D systems \citep{Servidio15,Li16}. From recent works, it has been suggested that 2D and 3D are in quantitative agreement \citep{Franci18}.

\begin{acks}
This project has received funding from the European Union’s Horizon 2020 research and innovation programme under grant agreement No 776262 (AIDA). This work is partly supported by the International Space Science Institute (ISSI) in the framework of International Team 405 entitled ‘Current Sheets,Turbulence, Structures and Particle Acceleration in the Heliosphere’, by the US NSF AGS-1156094 (SHINE). We acknowledge PRACE for awarding us access to SuperMUC at GCS@LRZ, Germany. The work by F. Pucci has been supported by Fonds Wetenschappelijk Onderzoek - Vlaanderen (FWO) through the postdoctoral fellowship 12X0319N. We thank the anonymous Referee for the comments and suggestions that helped improve the quality of this work. Disclosure of Potential Conflicts of Interest The authors declare that they have no conflicts of interest.
\end{acks}

\appendix

\begin{figure}[ht]
	\centering
  	{\includegraphics[width=.56\textwidth]{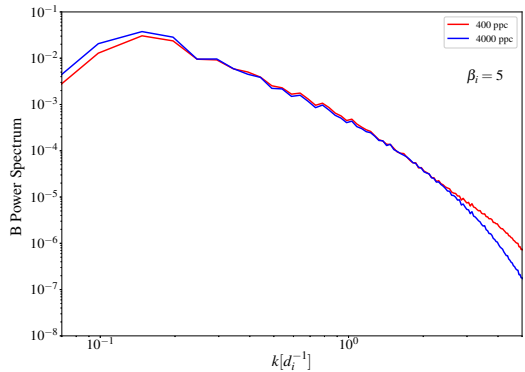}}
   \vspace{0.5cm}
   \caption{Magnetic field power spectra for run7 and run8 (K3) with 400 and 4000 particles per cell respectively. The spectra show the statistical converged achieved already at 400 ppc since the large scales are quite similar and the inertial ranges have the same slopes.}
   \label{fig:A1}
\end{figure}

The statistical convergence study for the hybrid runs has been performed in \cite{Servidio16}. They show that for a number of particles per cell larger than 400, turbulence statistical properties remain unchanged. Here we report the statistical convergence study performed for the full PIC high $\beta$ runs that are those more sensible to statistical noise. We verified the statistical convergence for both the power spectra and the diffusion coefficient. We found that the properties of turbulence remain unchanged for ppc $ \geq 400$. In \fig{fig:A1} we show the power spectra of the magnetic field for run7 (ppc $=400$) and run8 (K3) (ppc $=4000$). The spectra have the same behaviour (and inertial range slope) in the two cases. There are small differences, especially at very small (electron) scales, as expected.
To test the reliability of the high $\beta$ simulations also from the particle point of view, we compared the diffusion coefficient for the full PIC runs. We performed the analysis described in \sez{sec:diff} and measured the diffusion coefficient for each energy interval for the three different runs with $\beta_i=5$. The values of particle energy show consistency from one run to another meaning that acceleration and energization mechanisms are not affected by the number of particles per cell used (once convergence has been achieved). Also, the diffusion coefficients are comparable among the runs (at the same energy) meaning that neither spatial diffusion is affected by variation in the number of particles used, further proving that convergence has been achieved for this high-$\beta$ case. This agreement is somehow expected since the diffusion coefficient depends on the shape of the spectrum in the energy-containing and inertial scales $(k d_i << 1)$ that are similar as shown in \fig{fig:A1}.

\clearpage
\newpage

\begin{table}
\begin{tabular}{cc|cc|cc}
    \hline
    \multicolumn{2}{c}{run7} & 
    \multicolumn{2}{c}{run8 (K3)} & 
    \multicolumn{2}{c}{run9} \\
    \hline
$v_z^2$ & $D$ & $v_z^2$ & $D$ & $v_z^2$ & $D$ \\ 
\hline 
0.6 & 2.0 & 0.7 & 2.1 & 0.6 & 2.1 \\ 
\hline 
1.9 & 4.0 & 2.0 & 4.3 & 1.8 & 4.0 \\ 
\hline 
3.1 & 5.4 & 3.3 & 6.0 & 3.1 & 5.3 \\ 
\hline 
4.3 & 6.3 & 4.6 & 7.1 & 4.3 & 6.5 \\ 
\hline 
5.6 & 7.1 & 5.9 & 7.7 & 5.5 & 7.3 \\ 
\hline 
\end{tabular}
\caption{Particle energy values and the corresponding diffusion coefficient. The energy values do not change by increasing the number of particles from 400 to 4000 meaning that particles are accelerated and energized in the same way. Moreover, the diffusion coefficients are comparable among the runs suggesting that also spatial diffusion is not affected by an increase in the number of particles.}
\label{tab:tab2}
\end{table}

\bibliographystyle{plainnat}
\bibliography{biblio}

\end{article}

\end{document}